\documentclass[twocolumn]{aastex63}

\usepackage{bm}

\newcommand{\unitlum}{\rm{erg}~{\rm s}^{-1}}
\newcommand{\unitnum}{\rm{cm}^{-3}}
\newcommand{\unitb}{\rm{erg}~\rm{cm}^{-3}}

\newcommand{\unitK}{\rm{K}}

\newcommand{\nh}{n_{\rm H}}

\newcommand{\zd}{Z_{\rm d}}
\newcommand{\tg}{T_{\rm g}}
\newcommand{\td}{T_{\rm d}}
\newcommand{\tsub}{T_{\rm sub}}

\newcommand{\jpe}{J_{\rm pe}}
\newcommand{\jion}{J_{\rm ion}}
\newcommand{\je}{J_{\rm e}}
\newcommand{\jsec}{J_{\rm sec,gas}}
\newcommand{\qabs}{Q_{\rm abs}}

\newcommand{\smax}{S_{\rm max}}
\newcommand{\hsmax}{\hat{S}_{\rm max,9}}
\newcommand{\lang}{L_{\rm AGN}}
\newcommand{\uion}{U_{\rm ion}}
\newcommand{\asub}{a_{\rm sub}}
\newcommand{\ace}{a_{\rm CE}}
\newcommand{\amax}{a_{\rm max}}
\newcommand{\amin}{a_{\rm min}}

\shorttitle{Dust destruction by charging in AGN}
\shortauthors{Tazaki, Ichikawa, and Kokubo}

\begin{document}

\title{Dust Destruction by Charging: A Possible Origin of Grey Extinction Curves of Active Galactic Nuclei}

\correspondingauthor{Ryo Tazaki}
\email{rtazaki@astr.tohoku.ac.jp}

\author[0000-0003-1451-6836]{Ryo Tazaki}
\affil{Astronomical Institute, Graduate School of Science
Tohoku University, 6-3 Aramaki, Aoba-ku, Sendai 980-8578, Japan}

\author[0000-0002-4377-903X]{Kohei Ichikawa}
\affil{Astronomical Institute, Graduate School of Science
Tohoku University, 6-3 Aramaki, Aoba-ku, Sendai 980-8578, Japan}
\affil{Frontier Research Institute for Interdisciplinary Sciences, Tohoku University, Sendai 980-8578, Japan}

\author[0000-0001-6402-1415]{Mitsuru Kokubo}
\affil{Astronomical Institute, Graduate School of Science
Tohoku University, 6-3 Aramaki, Aoba-ku, Sendai 980-8578, Japan}




\begin{abstract}
Observed extinction curves of active galactic nuclei (AGNs) are significantly different from those observed in the Milky Way. The observations require preferential removal of small grains at the AGN environment; however, the physics for this remains unclear. 
In this paper, we propose that dust destruction by charging, or Coulomb explosion, may be responsible for AGN extinction curves. Harsh AGN radiation makes a dust grain highly charged through photoelectric emission, and grain fission via Coulomb explosion occurs when the electrostatic tensile stress of a charge grain exceeds its tensile strength. We show that Coulomb explosion can preferentially remove both small silicate and graphite grains and successfully reproduce both flat extinction curve and the absence of 2175\AA~bump.
\end{abstract}



 \section{Introduction}\label{sec:intro}
Dust is a crucial component of active galactic nuclei (AGNs) \citep{urr95}. Recently, mid-infrared interferometric observations have revealed that the presence of dust grains at polar regions at pc-scales \citep{hon12,hon13,tri14,lop16,lef18,hon19}.
These polar dust grains are thought to be irradiated by harsh AGN radiation almost directly, and grain properties could be different from those observed in the local interstellar medium \citep[e.g.,][]{lao93}. 

The wavelength dependence of extinction at ultraviolet wavelengths is a powerful tool to infer dust properties at the polar region because grain properties are imprinted in the extinction curves \citep[e.g.,][]{Li07}.
Previous observations have shown that AGN extinction curves are significantly different from those observed in the Milky Way \citep{M01a, M01b}. Major properties of AGN extinction curves are (i) flat wavelength dependence at far-ultraviolet wavelengths and (ii) the absence of 2175\AA~bump \citep{G04, C04, G07}, which is thought to be caused by small graphite grains and/or polycyclic aromatic hydrocarbon (PAH) nanoparticles \citep[e.g.,][]{D84,WD01b,C11}. These observations imply that small grains, in particular for graphite grains, are preferentially removed from the AGN environments.

Several mechanisms have been proposed to explain the depletion of small grains, such as thermal sublimation and sputtering \citep{lao93}; however, these models seem to fail. The sublimation is more likely to remove silicate grains rather than graphite grains, which is not consistent with observations \citep[e.g.,][]{G04}. 
Chemisputtering might also preferentially destroy hot graphite grains \citep{Bar78,D79b}; however, it might be suppressed for highly charged grain as thermal sputtering is suppressed at the vicinity of AGN \citep[e.g.,][]{RT20}. 
In addition, drift-induced sputtering may not be an efficient mechanism for destroying small grains ($\lesssim0.1~\mu$m) because Coulomb coupling between gas and the grains tends to halt hypersonic drift \citep{RT20}. Although \citet{H19} pointed out rotational disruption recently, this mechanism is also inefficient for disrupting small grains. 
If 2175~\AA~bump is associated with PAH nanoparticles \citep[e.g.,][]{Li01}, these small grains might be disrupted by stochastic heating at around AGN, although this possibility is also inconclusive. Hence, up to date, a physical process responsible for flat and featureless extinction curves is still a matter of debate.

In this paper, we propose a new scenario for the origin of AGN extinction curves: dust destruction by charging, or Coulomb explosion \citep[e.g.,][]{D79, WDB06}. Dust destruction by charging has been discussed in the field of gamma-ray burst \citep{W00, F01}; however, such effect has been overlooked in interpreting AGN extinction curves. Since \citet{WDB06} have shown that Coulomb explosion can occur even if grains are 100 pc away from quasar, it is naturally anticipated that such process may significantly alters grain properties at pc-scale AGN environment.

In this paper, we study how dust destruction by charging affects AGN extinction curves and compare our model with previously suggested thermal sublimation model. 
This paper is organized as follows. In Section \ref{sec:model}, we summarize methods to calculate Coulomb explosion and thermal sublimation. 
Extinction curves predicted by dust destruction models are presented in Section \ref{sec:results}. 
Sections \ref{sec:disc} and \ref{sec:summary} present Discussion and Summary, respectively.

\begin{figure*}[htbp]
\begin{center}
\includegraphics[height=5.8cm,keepaspectratio]{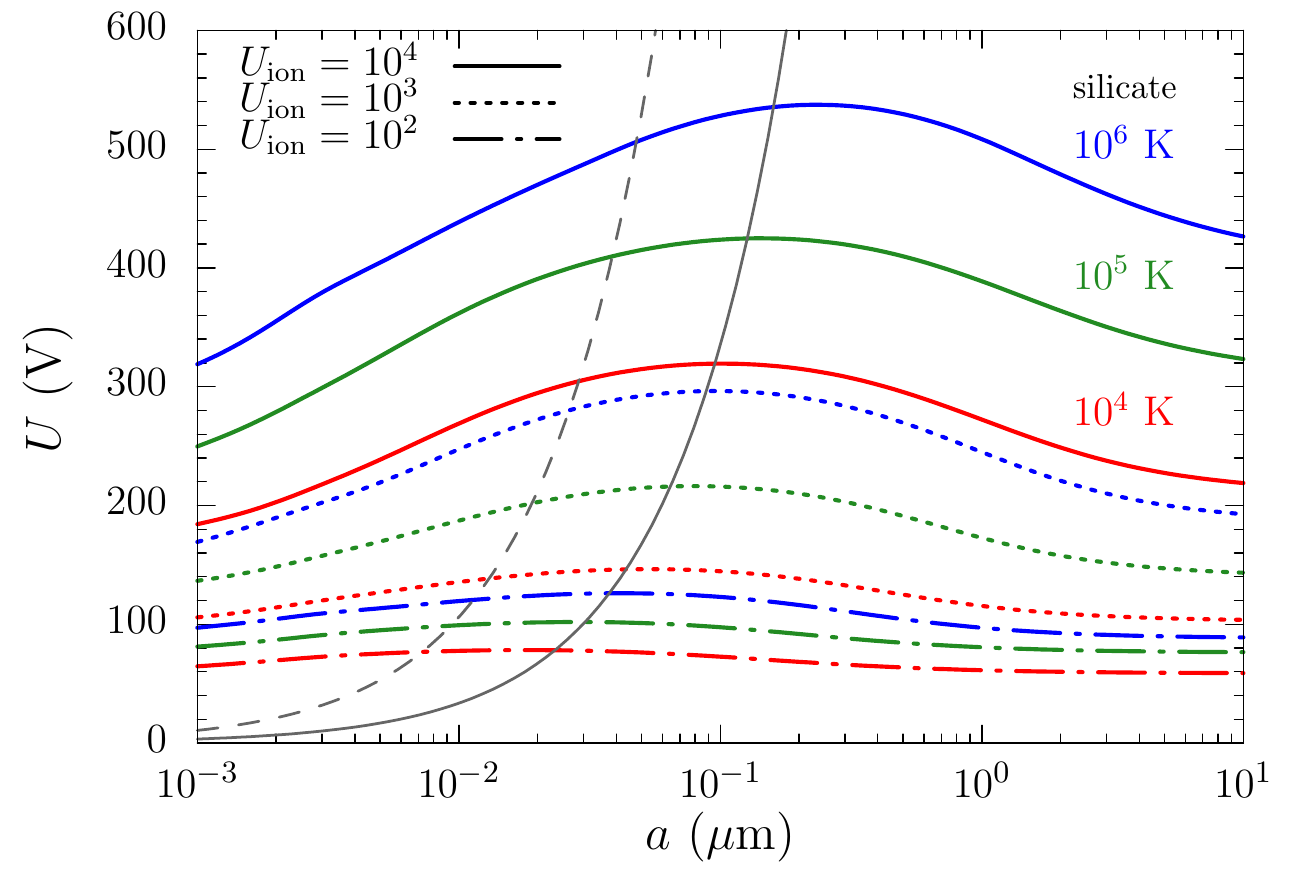}
\includegraphics[height=5.8cm,keepaspectratio]{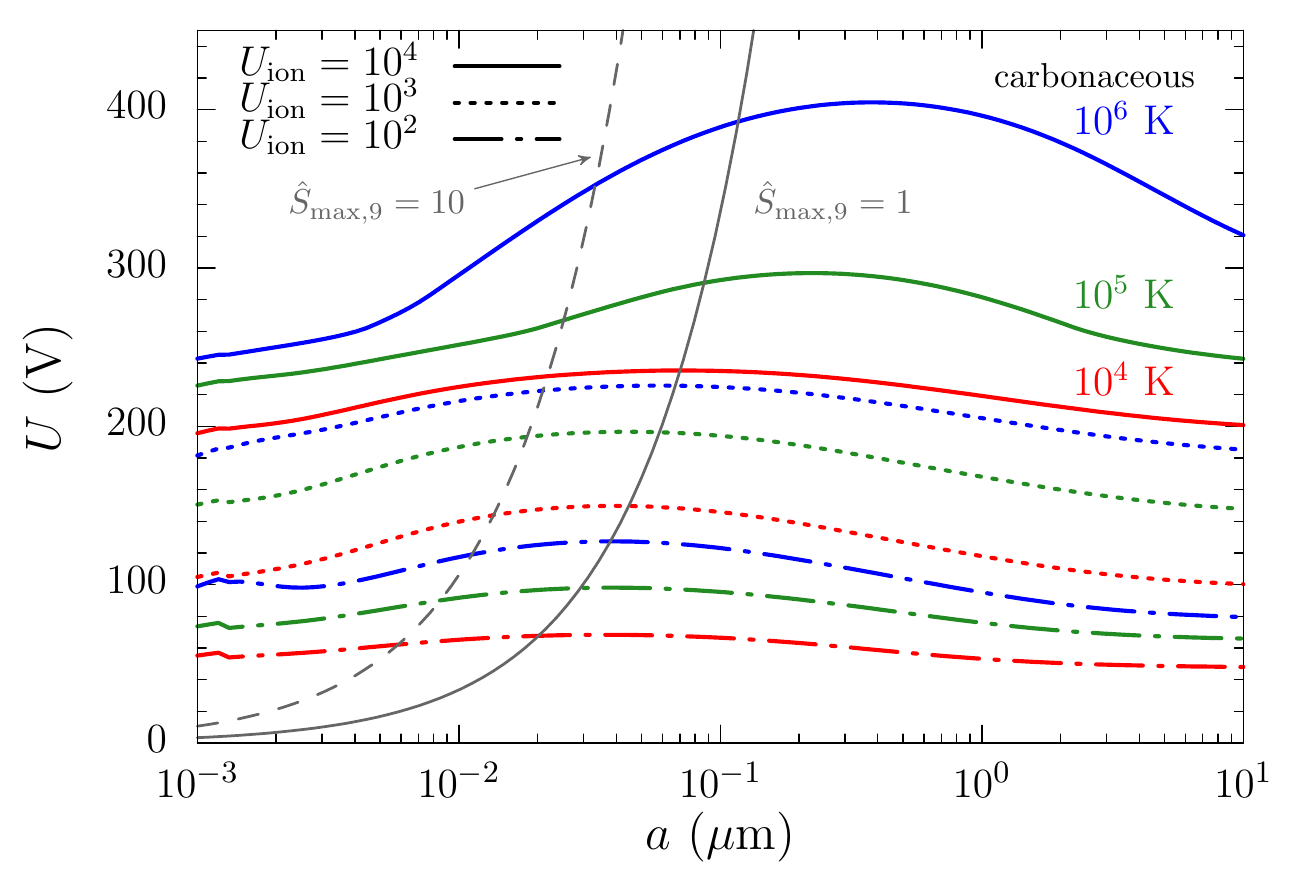}
\caption{The electrostatic potential, $U=eZ_\mathrm{d}/a$, for silicate grains (left) and carbonaceous grains (right). Solid, dotted, and dot-dashed lines represent the results for $U_\mathrm{ion}=10^4$, $10^3$, and $10^2$, respectively. Blue, green, and red colors represent gas temperature $\tg=10^{6}$ K, $10^{5}$ K, and $10^{4}$ K, respectively. Gray solid and dashed lines represent the threshold for Coulomb explosion for grains of the tensile strength $\hsmax=1$ and $10$, respectively.}
\label{fig:charge}
\end{center}
\end{figure*}

\section{Methods and Models} \label{sec:model}
\subsection{AGN Environment}
The radiation spectra of AGNs are taken from \citet{N08} and the bolometric luminosity is assumed to be $\lang=10^{45}~\unitlum$. For convenience, we define $L_{45}=(\lang/10^{45}~\unitlum)$.
Since we focus on grains at polar region, where grains are thought to be irradiated by AGN radiation directly, we ignore attenuation of AGN radiation. 
At the pc-scale polar region, radiation-hydrodynamic simulations suggests that the gas temperature and density are about $\tg\approx10^4$ K and $\nh\approx10-10^3~\unitnum$, respectively \citep{W16}. Hence, we adopt $\tg=10^4$ K and $\nh=10^2~\unitnum$ as a fiducial set of parameters. For the sake of simplicity, we assume $\tg$ and $\nh$ are constants.

\subsection{Grain Charge}  \label{sec:graincharge}
Dust grains become positively charged in the AGN environments \citep{WDB06}.
We compute grain charge $\zd$ (in the electron charge unit) by solving the rate equation \citep[][]{WDB06, RT20}:
\begin{equation}
\frac{d\zd}{dt}=\jpe-\je+\jsec+\jion, \label{eq:charge}
\end{equation}
where $\jpe$ is the photoelectric emission rate \citep{WD01, WDB06}, $\je$ and $\jion$ are the collisional charging rate of electrons and ions, respectively \citep{D87}, and $\jsec$ is the rate for secondary electron emission induced by incident gas-phase electrons \citep{D79}. 
Typical charging timescale of a neutral grain due to electron collisions is $\je^{-1}\sim0.9$ s \citep{D87}, where $\tg=10^4$ K, grain radius $a=0.1~\mu$m, the electron density $n_e=10^2~\mathrm{cm}^{-3}$, and the sticking probability of 0.5 \citep{D87} are used.
Since charging timescale is much shorter than dynamical timescale, we can assume the steady state in Equation (\ref{eq:charge}). 
In addition, we can also ignore grain charge distribution because the single-charge equilibrium approximation gives reliable results for highly charged grains \citep{WDB06}. Thus, in this paper, we solve $\jpe-\je+\jsec+\jion=0$ to find $\zd$.

It is useful to introduce the ionization parameter, $U_\mathrm{ion}\equiv n_\gamma/n_\mathrm{H}$, where $n_\gamma$ is the total photon number density beyond 13.6 eV. 
Since the grain charge is mostly determined by the balance of photoelectric emission and electron collisions, $U_\mathrm{ion}$ characterizes the grain-charge amount.
For the radiation spectra of AGNs used in \citet{N08}, we obtain
\begin{eqnarray}
n_\gamma&=&2.61\times10^{6}~\mathrm{cm}^{-3}\left(\frac{L_\mathrm{AGN}}{10^{45}~\mathrm{erg}~\mathrm{s}^{-1}}\right)\left(\frac{r}{\mathrm{pc}}\right)^{-2},
\end{eqnarray}
where $r$ is the distance from the central engine of AGN.
As a result, the ionization parameter becomes
\begin{eqnarray}
U_\mathrm{ion}=2.61\times10^{4}\left(\frac{n_\mathrm{H}}{10^2~\mathrm{cm}^{-3}}\right)^{-1}\left(\frac{L_\mathrm{AGN}}{10^{45}~\mathrm{erg}~\mathrm{s}^{-1}}\right)\left(\frac{r}{\mathrm{pc}}\right)^{-2}. \label{eq:uion}
\end{eqnarray}

\subsection{Grain Temperature}
The grain temperature, $\td$, is obtained from radiative equilibrium:
\begin{equation}
\frac{\lang}{4\pi r^2}\pi a^2 \langle \qabs \rangle_\mathrm{AGN}=4\pi a^2 \sigma_\mathrm{SB}
\td^4  \langle \qabs \rangle_{\td}, \label{eq:radeq}
\end{equation}
where $r$ is the distance from the AGN, $\langle \qabs \rangle_\mathrm{AGN}$ and $\langle \qabs \rangle_{\td}$ are AGN-spectrum averaged absorption efficiency and Planck mean absorption efficiency at dust temperature $\td$, respectively, and $\sigma_\mathrm{SB}$ is the Stefan--Boltzmann constant.
The absorption efficiency, $\qabs$, is calculated by using the Mie theory \citep{B83}, and the optical constant of silicate grains was taken from \citet{D84, lao93, D03}. For graphite, the optical constant is calculated by adding the interband and free electron contributions \citep[see also][]{D84, lao93, D03}, where we adopt free electron models of \citet{A12}. If the size parameter $x=2\pi{a}/\lambda$, where $\lambda$ is the wavelength, is larger than $2\times10^{4}$, we use the anomalous diffraction approximation \citep{VDH57} instead of using the Mie theory. 

\subsection{Grain Destruction Processes}
We consider two kinds of dust destruction: Coulomb explosion (Section \ref{sec:CE}) and thermal sublimation (Section \ref{sec:TS}).

\subsubsection{Coulomb explosion} \label{sec:CE}
If a dust grain acquires large amount of positive charges, Coulomb repulsion force within the grain causes grain fission, so-called Coulomb explosion \citep{D79}. The condition for Coulomb explosion is \citep{D79},
\begin{equation}
S=\frac{1}{4\pi}\left(\frac{U}{a}\right)^2 \ge \smax,
\end{equation}
where 
$S$ is the tensile stress in a charged sphere, $U=\zd{e}/a$ is the electrostatic grain potential, and $\smax$ is the tensile strength of the material. In the following, we use the normalized tensile strength defined by $\hsmax=(\smax/10^9~\unitb)$.

If the grain potential satisfies $U(a)\ge a(4\pi\smax)^{1/2}$, the electric stress exceeds tensile strength of a grain, and then, Coulomb explosion will occur. We define the critical grain radius for Coulomb explosion, $\ace$, such that $U(\ace)=\ace(4\pi\smax)^{1/2}$. Grains smaller than $\ace$ are subjected to Coulomb explosion.
Coulomb explosion will produce fragments of smaller grains. However, smaller fragments will be also charged enough to cause Coulomb explosion. Hence, we expect that a cascade fragmentation of grains due to Coulomb explosion occurs.

The tensile strength of a dust grain depends on material properties, such as composition and crystallinity.
Although the tensile strength of cosmic dust particles is highly uncertain, laboratory measurements gives us an estimate \citep[see][for a summary of tensile strength]{H19}. The tensile strength of graphite (polycrystalline) is suggested to be about $\hsmax=0.5 - 1$ \citep{B74}. 
Hence, in this study, we adopt the conservative value of $\hsmax=1$ for graphite grains. 
The tensile strength of forsterite (silicate) can be as small as $\hsmax=1.21$ \citep{G19}. Meanwhile, \citet{M72} reported the tensile strength of glass rods and fibers are about $\hsmax=130$. We adopt the values of $\hsmax=10$ for silicate.
It is worth to note that above measurements are based on bulk materials, and therefore, small grains may have different values of tensile strength.
Since the tensile strength is uncertain parameter, we discuss how $\smax$ changes our results in Section \ref{sec:results}. 

\subsubsection{Thermal sublimation} \label{sec:TS}
The sublimation temperature of dust grains, $\tsub$, is determined by a balance between gas pressure and saturation pressure \citep{G89, bas18}.
By using Equation (27) in \citet{bas18}, we compute $\tsub$ with the the standard solar elemental abundances \citep{Grevesse98}. As a result, we obtain $\tsub=1322~\unitK$ for graphite and $\tsub=1072~\unitK$ for silicate grains when $\nh=10^2~\unitnum$ and $\tg=10^4~\unitK$.

Since smaller grains are usually hotter than larger grains, they can preferentially sublimate \citep[e.g.,][]{lao93}. We can define critical grain radius for thermal sublimation such that $\td(\asub)=\tsub$. Grains smaller than $\asub$ will sublimate.

\section{Results} \label{sec:results}
\subsection{Electrostatic Grain Potential} \label{sec:charge}
We first solve Equation (\ref{eq:charge}) with the steady state assumption, and the results are presented in Figure \ref{fig:charge}. As a general tendency,  higher $\uion$ and $\tg$ gives larger grain electrostatic potential. Our calculations are quantitatively agree with \citet{WDB06}, although assumed radiation spectra are not the same. 

Figure \ref{fig:charge} shows that the grain potential depends on the grain radius and becomes maximum at the sub-micron sizes, while grain charge $\zd$ is a monotonically increasing function of grain radius.
This is mainly determined by two competing effects: photoelectric yield and photon absorption efficiency. With decreasing grain radius, the photoelectric yield is increased due to the small particle effect \citep[e.g.,][]{W73, D78}. Hence, smaller grains are more likely to emit photoelectrons once a high energy photon is absorbed. However, decreasing grain radius reduces the absorption efficiency of photons, once the grain radius is smaller than the incident radiation wavelength, which is typically about $\lambda\sim0.1$~$\mu\mathrm{m}$. Hence, due to lower photon absorption efficiency, the photoelectric emission rate is decreased, and therefore, the grain potential is decreased for smaller grains ($a\lesssim0.1~\mu\mathrm{m}$). 
Because of these two effects, the grain potential is maximized at sub-micron sizes. 
Although the grain potential also depends on the grain composition, silicate and graphite grains have almost similar grain potential.

Figure \ref{fig:charge} also shows that the critical grain size for Coulomb explosion is about $\ace\sim0.01-0.1~\mu$m for $\uion=10^2-10^4$ and $\tg=10^4-10^6$ K at $\hsmax=1$. 

\begin{figure}[t]
\begin{center}
\includegraphics[height=6.0cm,keepaspectratio]{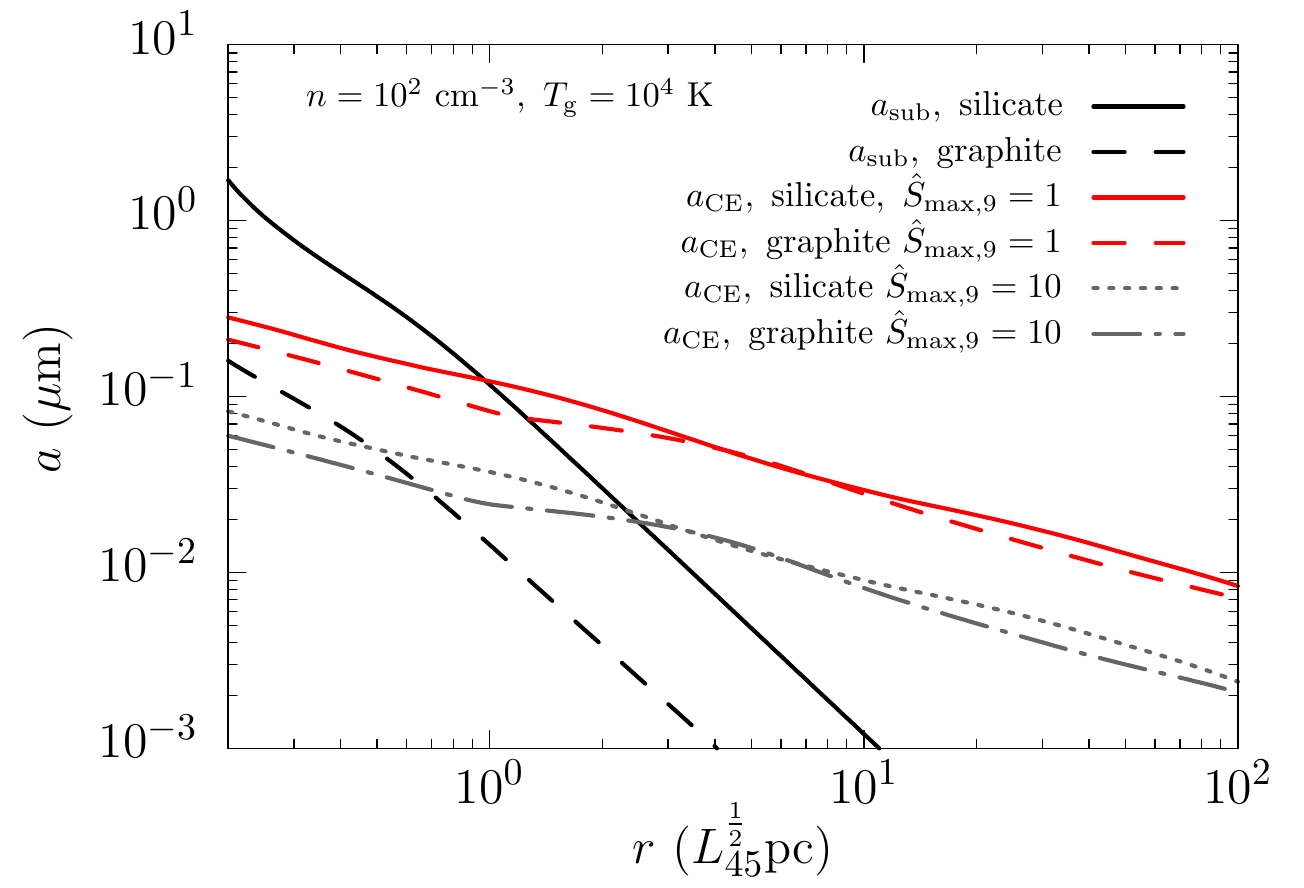}
\caption{Grain radius against a distance from AGN. 
Black and red lines indicate $\asub$ and $\ace$, respectively, where $\ace$ is computed for $\hsmax=1$ for both silicate and graphite grains. Solid and dashed lines indicate silicate and graphite grains, respectively. Grey dotted and dash-dotted lines are $\ace$ for $\hsmax=10$ for both graphite and silicate grains, respectively. Grains with $a\le\asub,~\ace$ will be disrupted.}
\label{fig:size}
\end{center}
\end{figure}

\subsection{Coulomb Explosion versus Thermal Sublimation} \label{sec:vs}
Next, we compare $\ace$ and $\asub$ as well as their radial dependence from the center of AGN.
Figure \ref{fig:size} shows the critical grain radii $\asub$ and $\ace$ for both silicate and graphite composition as a function of the distance from AGN.
It is found that $\ace$ has shallower radial dependence than $\asub$. 
The radial dependence of the critical grain radius for sublimation is about $\asub\propto r^{-2}$. As long as the Rayleigh approximation is valid, that is, wavelength of thermal emission is longer than the grain radius, we have $\langle \qabs \rangle_{\td}\propto a$. Hence, for a fixed dust temperature ($\td=\tsub$), Equation (\ref{eq:radeq}) results in $\asub\propto r^{-2}$. In other words, $\asub$ is proportional to the radiative flux from AGN. 

\begin{figure}[t]
\begin{center}
\includegraphics[height=7.5cm,keepaspectratio]{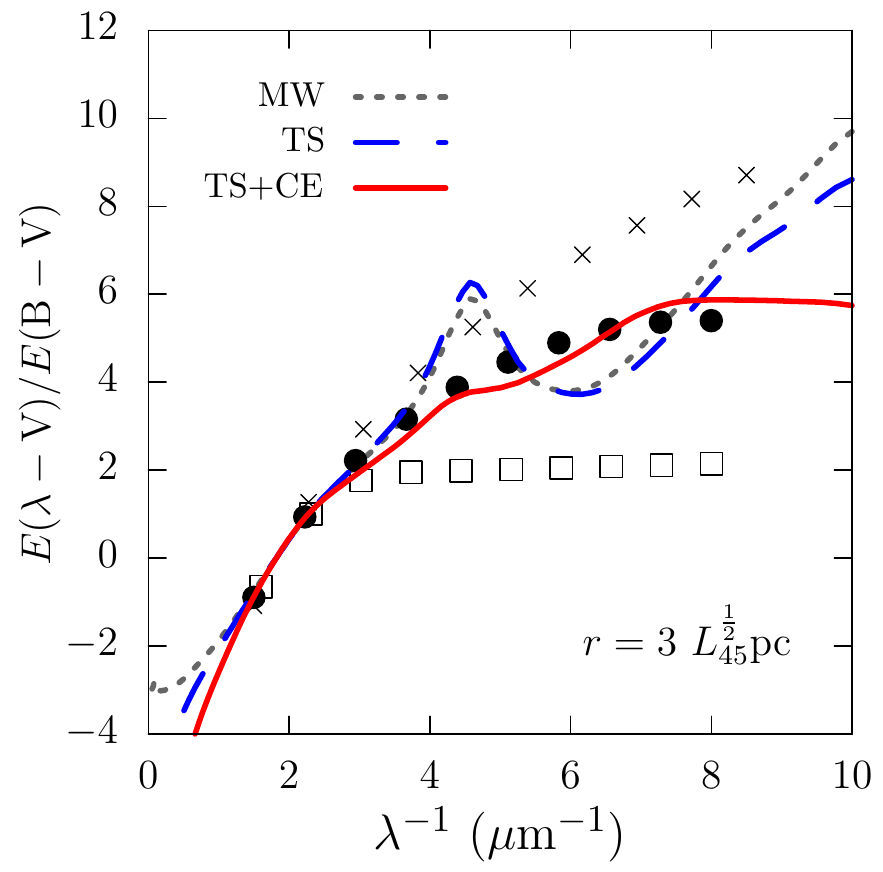}
\caption{Extinction curves with and without Coulomb explosion are shown in red-solid and blue-dashed lines, respectively. 
For reference, grey short-dashed line is the average Milky Way extinction curve, which is computed with $\amin=0.005~\mu$m and $\amax=0.25$~$\mu$m.
Circles indicate observed values taken from \citet{G07}, whereas crosses and squares indicate those taken from \citet{C04} and \citet{G04}, respectively. We assume $\amax=1~\mu\mathrm{m}$.}
\label{fig:ext}
\end{center}
\end{figure}

\begin{figure*}[t]
\begin{center}
\includegraphics[height=9cm,keepaspectratio]{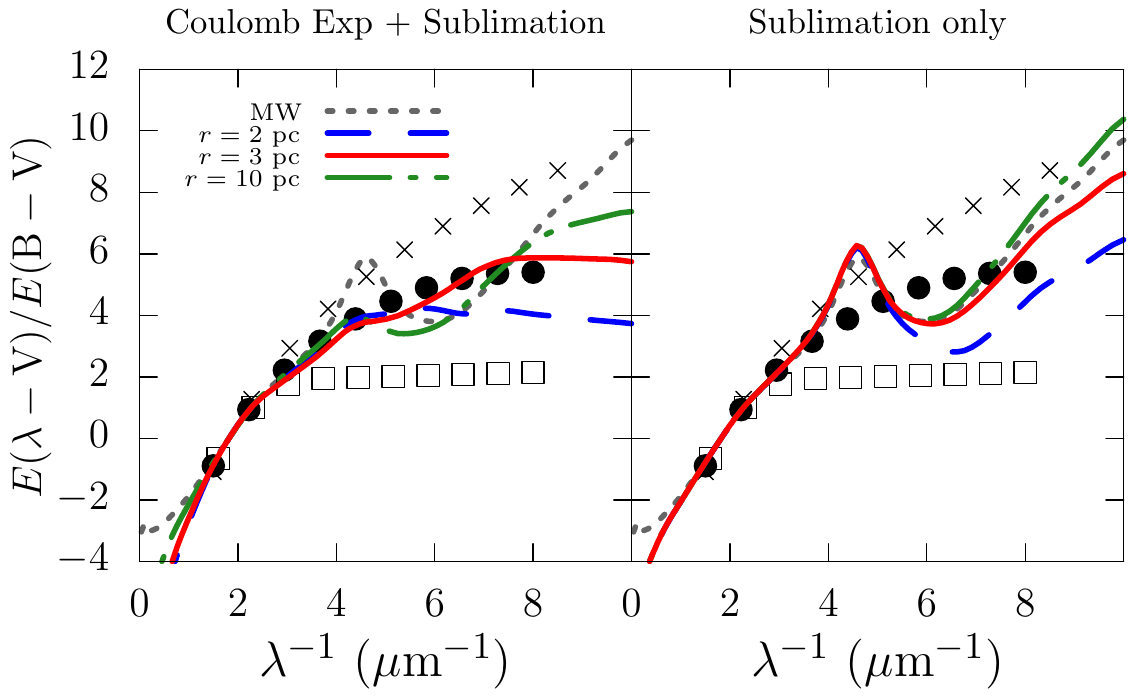}
\caption{Same as Figure \ref{fig:ext}, but for different distances from AGN. For all lines, the AGN luminosity is set as $L_{45}=1$. Left and right panels show the results for with and without Coulomb explosion, respectively.}
\label{fig:ext2}
\end{center}
\end{figure*}

Grain charging seems to be caused also by the radiation flux, since a grain is charged via photoelectric emission.
However, grain potential is {\it not} proportional to $r^{-2}$. For example, in Figure \ref{fig:charge}, even if $\uion$ ($\propto$ radiative flux) decreases by an order of magnitude, grain potential $U$ decreases only by a factor of few. 
This results suggest that Coulomb explosion can be important at larger distances, e.g., pc-scale distance.

Figure \ref{fig:size} also shows that a difference in $\ace$ between graphite and silicate grains is not so large compared to the difference seen in $\asub$. Thus, Coulomb explosion tends to remove both silicate and graphite grains of almost similar grain radii.
Meanwhile, for sublimation, the graphite grains have smaller $\asub$ because (1) the emissivity at near-infrared wavelength is higher, and (2) $\tsub$ is higher \citep[see also][]{lao93, bas18}.
Therefore, this suggests that thermal sublimation preferentially remove small silicate grains rather than graphite grains.

\subsection{Extinction Curves} \label{sec:ext}
To understand how thermal sublimation or Coulomb explosion changes extinction curves, we compute extinction cross section of grains.
We assume that the grain size distribution obeys $dn_i\propto \mathcal{A}_ia^{-3.5}da$ ($\amin^i \le a \le \amax$), where $dn_i$ is the number density of dust grains of type $i$ (silicate or graphite) in a size range $[a,a+da]$, and $\mathcal{A}_i$ is the abundance of the grain type $i$ \citep{D84}.
We set $\amin^i=\mathrm{max}(\asub^i$,$\ace^i$,$a_\mathrm{min,MRN}$), 
where $a_\mathrm{min,MRN}=0.005~\mu$m.
We treat $\amax$ as a free parameter. We also set the abundance of silicate and graphite from \citet{D84}. 
The extinction magnitude at wavelength $\lambda$, $A_\lambda$, is
\begin{equation}
    A_\lambda\propto \sum_i \int C_\mathrm{ext}^i(\lambda,a)\frac{dn_i}{da}da,
\end{equation}
where $C_\mathrm{ext}^i(\lambda,a)$ is the extinction cross section of a grain of type $i$ with radius $a$. We define the extinction curve as $E(\lambda-V)/E(B-V)=(A_\lambda-A_V)/(A_B-A_V)$, where $A_V$ and $A_B$ are the extinctions at visual (5500~\AA) and blue (4400~\AA) wavelengths.

Figure \ref{fig:ext} shows the extinction curves for with/without Coulomb explosion at $3 L_{45}^{\frac{1}{2}}$ pc away from the nucleus.
The extinction curve with Coulomb explosion can successfully reproduce flat extinction curve as well as the absence of $2175$~\AA~bump. In addition, predicted extinction curve is consistent with the observation by \citet{G07}.
Meanwhile, the extinction curve without Coulomb explosion, or $\amin^i=\mathrm{max}(\asub^i,a_\mathrm{min,MRN})$, shows prominent $2175$~\AA~bump. This is because sublimation does not remove small graphite grains, although small silicate grains are removed (Figure \ref{fig:size}). 
Since observed AGN extinction curves do not show such a bump \citep{G04, C04, G07}, the sublimation model is insufficient to reproduce observed extinction curves.

\begin{figure}[t]
\begin{center}
\includegraphics[height=7.5cm,keepaspectratio]{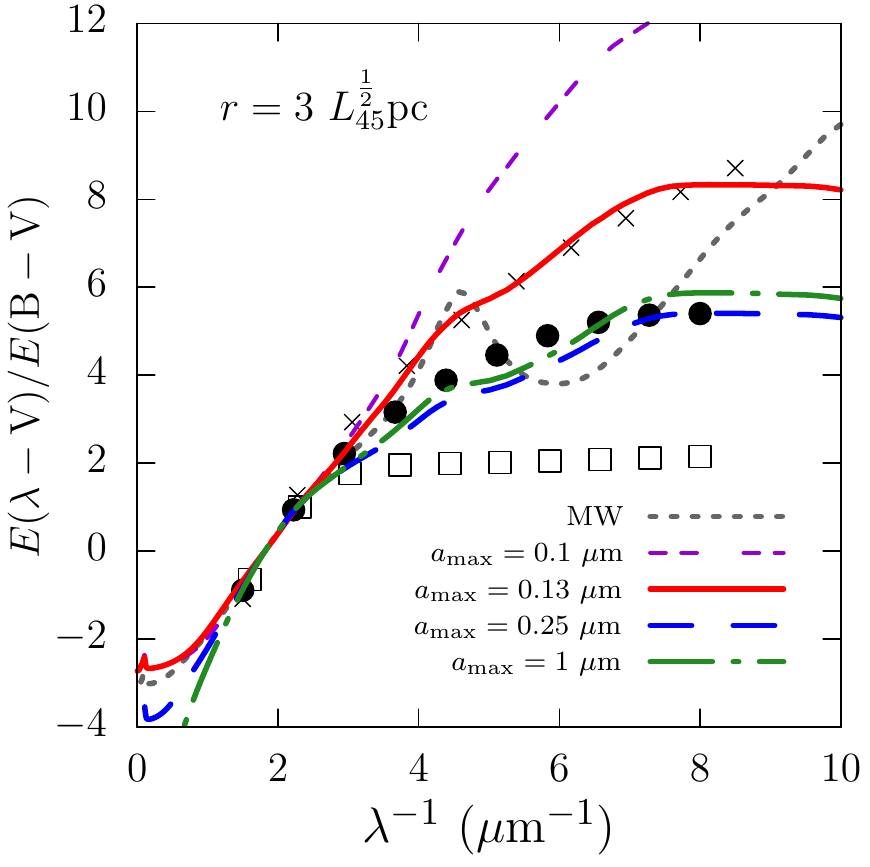}
\caption{Extinction curves for various $\amax$, where tensile strength of graphite and silicate grains are $\hsmax=1$ and $10$, respectively. 
$r=3 L_{45}^{\frac{1}{2}}$ pc is assumed. Grey short-dashed line shows the Milky-way values.}
\label{fig:ext3}
\end{center}
\end{figure}

\begin{figure}[t]
\begin{center}
\includegraphics[height=7.5cm,keepaspectratio]{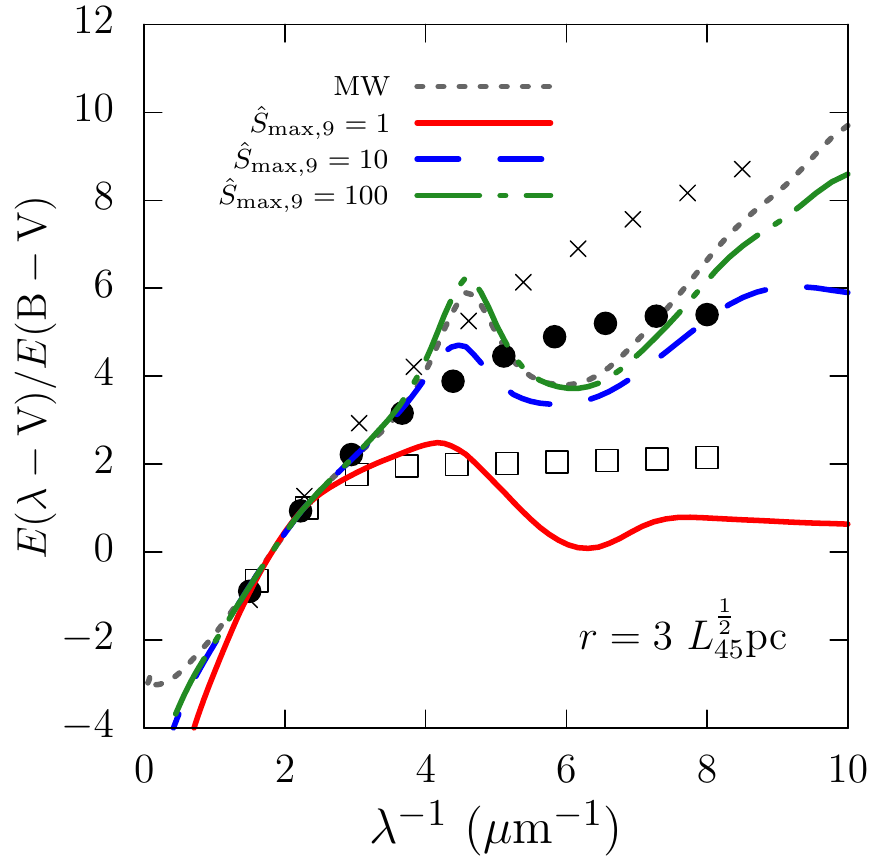}
\caption{The predicted extinction curves by the Coulomb explosion model with various tensile strength values. Solid, dashed, and dot-dashed lines correspond to $\hsmax=1,~10,~100$, respectively. Both silicate and graphite grains have the same tensile strength with assumption of $\amax=1~\mu$m. }
\label{fig:ext4}
\end{center}
\end{figure}

Figure \ref{fig:ext2} shows extinction curves at various radial distances.
Even if the radial distances are changed, overall shape of the extinction curve with Coulomb explosion is still similar with the one from \citet{G07} up to $r\simeq10 L_{45}^{\frac{1}{2}}$~pc, while the thermal sublimation model remains inconsistent with observations.
With increasing the distance from AGN, $\uion$ decreases, and then, smaller grains can survive from Coulomb explosion; nevertheless, $2175$~\AA~is still weak for the model with Coulomb explosion.

Extinction curves with Coulomb explosion are not sensitive to the choice of maximum grain radius as long as it is larger than 0.25$~\mu$m (Figure \ref{fig:ext3}). When grain radius is smaller this value, the extinction curve steeply increases with decreasing wavelength. 

Extinction curves with Coulomb explosion are found to be similar to those observed by \citet{G07}. In addition, if maximum grain radius is smaller than 0.25$~\mu$m, the curves become similar to those observed by \citet{C04}. However, our model fails to explain observations by \citet{G04}. Explaining such extinction curves might require additional mechanism, such as reduced-graphite abundance \citep{G04}.

\section{Discussion} \label{sec:disc}
\subsection{Comparison with observations of torus innermost radius}
We have shown that the Coulomb explosion can be more important process for dust destruction than thermal sublimation. Meanwhile, near-IR (NIR) dust reverberation mapping observations for AGNs, which show that the color temperatures of the variable hot dust emission agree with the dust sublimation temperature ($1400 - 2000$ K), and the innermost radius of the dust torus $R_{\mathrm{in}}$ is proportional to the square root of the AGN luminosity ($R_{\mathrm{in}} \propto \lang^{1/2}$), strongly suggesting that the dust innermost radius is determined by the thermal sublimation \citep[][and references therein]{yos14,kos14,mine19,gra20}. This apparent conflict can be attributable to the difference of gas density.

While the ambient gas density of $n_{\text{H}}=10^{2}~\text{cm}^{-3}$ assumed throughout the calculations in this work is a reasonable assumption for the outflowing gas at the polar region of AGNs, the gas density of the equatorial dust torus is expected to be much higher.
The gas at the innermost part of the dust torus can be as dense as broad emission line regions, where $\nh \sim 10^{10}~\unitnum$ \citep[e.g.,][]{bas18,kok20}; in such a dense gas region, higher dust sublimation temperatures are expected ($T_{\text{sub}}=1880$~K and $1503$~K for graphite and silicate grains, respectively), and the relative importance of the Coulomb explosion relative to the thermal sublimation is significantly reduced due to inefficient grain charging by the enhanced electron collision rate (see Section~\ref{sec:graincharge}).
Therefore, unlike in the case of the polar dust region, the dust destruction at the innermost region of the equatorial dust torus must be governed by the thermal sublimation.

\subsection{Tensile strength of small grains?}
Coulomb explosion depends on the tensile strength assumed. Figure \ref{fig:ext4}  shows how the tensile strength affects extinction curves. In Figure \ref{fig:ext4}, both silicate and graphite grains are assumed to have the same tensile strength. 
As the tensile strength increases, Coulomb explosion becomes inefficient, and then, both small silicate and graphite grains can survive. As a result, extinction curves shows an increase toward shorter wavelength with a prominent 2175~\AA~bump. 
Hence, observed extinction curves, e.g., a lack of 2175~\AA~bump, seems to be reproduced when $\hsmax\lesssim10$, and this is within a range of measured values for bulk materials (e.g., Section \ref{sec:CE}). 

Under the assumption of the tensile strength for bulk materials determined by laboratory experiments, the Coulomb explosion leads to the absence of small dust grains in the close vicinity of AGNs and thus can naturally explain the observed flat extinction curve. Conversely, if the tensile strength of the cosmic dust is far stronger than that for the bulk materials due to, e.g., crystallization by annealing for hot small grains, the graphite grains survive even under the large electrostatic potential and the 2175\AA\ bump feature is unavoidable. Therefore, if our scenario for the flat extinction curve by the Coulomb explosion is true, it also suggests that the tensile strength of the cosmic dust must be close to the value of the bulk materials. However, we should keep in mind that if PAH nanoparticles are the carrier of the 2175\AA\ bump, the bump might be suppressed by destroying these particles as suggested by observations \citep[e.g.,][]{Sturm00}.

\section{Conclusion}  \label{sec:summary}
In this paper, we have shown that thermal sublimation is insufficient to reproduce observed AGN extinction curves because this model predicts too strong 2175~\AA~bump due to preferential survival of small graphite grains. We have proposed that Coulomb explosion can successfully reproduces flat extinction curves as well as the absence of 2175~\AA~bump as long as the tensile strength is lower than $10^{10}~\unitb$. 
The predicted extinction curves have shown to be very similar to those observed by \citet{G07} as well as \citet{C04}.
The Coulomb explosion model implies that variety of radiation environment ($\uion$) and maximum grain radius may partly explain various types of observed AGN extinction curves \citep{C04, G07}. 

\acknowledgments
R.T. would like to thank Joseph Weingartner for his kind cooperation of our code validation. R.T. and M.K. were supported by a Research Fellowship for Young Scientists from the Japan Society for the Promotion of Science (JSPS) (JP17J02411 and  JP17J01884, respectively). This work was also supported by JSPS KAKENHI Grant Numbers JP19H05068 (R.T.). and JP18K13584 (K.I.). This work was supported by the Program for Establishing a Consortium
for the Development of Human Resources in Science
and Technology, Japan Science and Technology Agency (JST).

\bibliography{cite}
\end{document}